\documentclass[]{spie}  

\usepackage{amsmath,amsfonts,amssymb}
\usepackage{graphicx}
\usepackage[colorlinks=true, allcolors=blue]{hyperref}

\usepackage{tikz}
\usetikzlibrary{quantikz}

\title{Quantum Computing Calculations for Nuclear Structure and Nuclear Data}

\author[a]{Isaac Hobday}
\author[a]{Paul D. Stevenson}
\author[b]{James Benstead}
\affil[a]{Department of Physics, University of Surrey, Guildford, Surrey, GU2 7XH, UK}
\affil[b]{AWE, AWE, Aldermaston, Reading, RG7 4PR, UK}

\authorinfo{Further author information: (Send correspondence to P.D.S.)\\P.D.S.: E-mail: p.stevenson@surrey.ac.uk}

\begin{document} 
\maketitle

\begin{abstract}
Model calculations of nuclear properties are peformed using quantum computing algorithms on simulated and real quantum computers. 

The models are a realistic calculation of deuteron binding based on effective field theory, and a simplified two-level version of the nuclear shell model known as the Lipkin-Meshkov-Glick model.

A method of reducing the number of qubits needed for practical calculation is presented, the reduction being with respect to the number needed when the standard Jordan-Wigner encoding is used.  Its efficacy is shown in the case of the deuteron binding and shell model.  

A version of the variational quantum eigensolver in which all eigenstates in a spectrum are targetted on an equal basis is shown.  The method involves finding the minima of the variance of the Hamiltonian.  The method's ability to find the full spectrum of the simplified shell model is presented. 
\end{abstract}


\section{INTRODUCTION}
\label{sec:intro}  
Quantum computers are originally hypothetical and now realised devices which harness features of quantum mechanics to store and process data.  In a common design of quantum computer the basic processing unit is a \textit{qubit} -- a two-state system, commonly realised as a spin half fermion whose spin orientation gives the two states.  The possibility to simulate other quantum systems of spin half fermions through qubit operations was one of the earliest proposed uses of quantum computers \cite{feynman_simulating_1982}.  Subsequently, considerable work has taken place in developing and implementing quantum computing algorithms in applied quantum mechanics, most notably in quantum chemistry \cite{cao_quantum_2019,mcardle_quantum_2020} and in condensed matter systems \cite{bassman_simulating_2021}.

Nuclear physics is the study of systems of interacting protons and neutrons, both of which are spin half fermions.  As such, it is an area in which quantum computers can be applied.  Individual nuclei display a rich set of properties such as their binding energy, size, shape, spin, and their complex excited state spectra.  The joint experimental and theoretical understanding of these properties is the field of nuclear structure (as opposed to nuclear reactions), and is what concerns us here.  The starting point of theoretical studies of nuclear structure is the many-body Schr\"odinger equation, using the nuclear interaction as the potential term.  Practical calculations involve developing methods to make the solution of the Schr\"odinger equation possible, or managing the complicated and non-perturbative strong nuclear interaction, or (usually) both in concert.  In the work presented here, it is a method of making a practical solution of the many-body Schr\"odinger equation which is explored on quantum computers.

While nuclear physics problems are relatively unexplored on quantum computers, a few investigations have been made.  Among these is the determination of the ground state energy of the deuteron using a one-body Hamiltonian derived from quantum field theory.  This calculation was made using the variational quantum eigensolver (VQE) on IBM cloud quantum computers\cite{dumitrescu_cloud_2018};  The VQE calculation of the ground state of the Lipkin-Meshkov-Glick (LMG)\cite{lipkin_validity_1965}, again using IBM quantum computers\cite{cervia_lipkin_2021}; a comparison of encoding methods for nuclear Hamiltonians in terms of Pauli matrices\cite{siwach_quantum_2021};  wave function preparation of relevance to nuclear states\cite{siwach_filtering_2021,lacroix_symmetry-assisted_2020,https://doi.org/10.48550/arxiv.2110.06098,robbins_benchmarking_2021}; and preparation of excited nuclear states\cite{roggero_preparation_2020}. Some other recent applications in nuclear physics, particularly at the high energy interface between nuclear and particle physics, are covered in a review by Zhang \textit{et al.} \cite{zhang_selected_2021}.

In this contribution we present results of the deuteron Hamiltonian previously used by Dumitrescu \textit{et al.}\cite{dumitrescu_cloud_2018} but with an alternative encoding onto a reduced number of qubits, and we present an adapted VQE algorithm which allows ground and excited states to be found on an equal basis.  We apply this VQE algorithm to the deuteron and to the LMG model.

\section{Reduced qubit encoding}\label{sec:encoding}
In constructing a Hamiltonian for the deuteron, Dumitrescu \textit{et al.}\cite{dumitrescu_cloud_2018} formulated the problem in a harmonic oscillator basis as
\begin{equation}\label{eq:deuteron}
H_{N}=\sum_{n, n^{\prime}=0}^{N-1}\left\langle n^{\prime}|(T+V)| n\right\rangle a_{n^{\prime}}^{\dagger} a_{n},
\end{equation}
where $n$ labels $s$-wave oscillator states representing the deuteron wave function, $a^\dagger_n$ and $a_n$ create or destroy a deuteron in the state $n$, and $T$ and $V$ are the kinetic and potential energies whose matrix elements are
\begin{equation}
    \begin{aligned}
\left\langle n^{\prime}|T| n\right\rangle=& \frac{\hbar \omega}{2}\left[(2 n+3 / 2) \delta_{n}^{n^{\prime}}-\sqrt{n(n+1 / 2)} \delta_{n}^{n^{\prime}+1}\right.\\
&\left.-\sqrt{(n+1)(n+3 / 2)} \delta_{n}^{n^{\prime}-1}\right] \\
\left\langle n^{\prime}|V| n\right\rangle=& V_{0} \delta_{n}^{0} \delta_{n}^{n^{\prime}}
\end{aligned}
\end{equation}
with $V_0$=-5.68658111 MeV.  In the previous work, Dumitrescu \textit{et al.} used the standard Jordan-Wigner (JW) encoding to map the creation and annihilation operators in (\ref{eq:deuteron}) onto Pauli spin matrices as needed for direct manipulation of the quantum computer qubits.  This encoding maps each basis state in the expansion in (\ref{eq:deuteron}) to one qubit; hence $N$ qubits are needed for the $H_N$ problem. 

A more compact encoding may be explored by noting that the size of the Hilbert space spanned by a tensor product of the spaces of $n$ qubits is $2^n$ while the size of the space of the Hamiltonian (\ref{eq:deuteron}) grows linearly with $N$.  Hence a single qubit can fully encode the $H_2$ Hamiltonian, while the JW encoding requires two qubits, and two qubits can encode $H_4$, as opposed to the four qubits under JW.

Writing out the Hamiltonian (\ref{eq:deuteron}) explicitly in matrix form in the oscillator basis and decomposing directly into tensor products of Pauli spin matrices\cite{rocco_h2zixy} gives 
\begin{equation}
    H_2 = 5.906709 I_0 - 6.343291 Z_0 - 4.286607 X_0.
\end{equation}
Here the operators are the set of Pauli operators including the identity, $\{I,X,Y,Z\}$, and the subsript $0$ refers to a qubit label.  In this case there is one qubit.  Note that the JW encoding requires two qubits and gives $H_{2,JW}=5.906709I + 0.218291 Z_0 - 6.125 Z_1 - 2.143304 (X_0X_1 + Y_0Y_1)$.

We use a hardware general ansatz for the wave function in the variational method, written in quantum circuit form as
			        
\centerline{\begin{quantikz}
\lstick{\ket{0}} & \qw & \gate{R_y(\theta_1)} & \qw & \gate{R_z(\theta_2)} & \qw .
\end{quantikz}}

\begin{figure}
    \centering
    \includegraphics[width=0.6\textwidth]{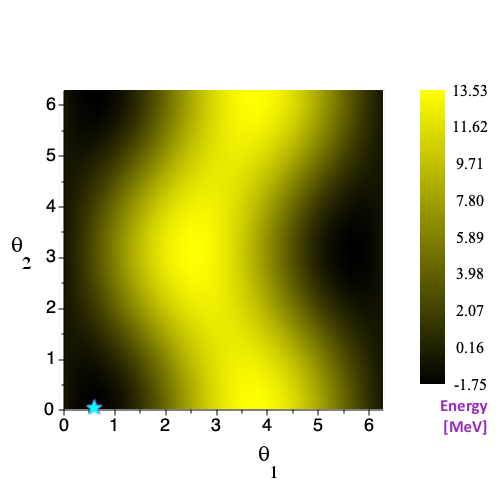}
    \caption{Expectation value of $H_2$ Hamiltonian with reduced qubit encoding as decribed in the text, with two-parameter general ansatz.  The cyan star shows the approximate minimum.}
    \label{fig:esurf}
\end{figure}

For a low-N problem like this we can get an idea of the energy landscape by explicitly mapping out the computed expectation value of the energy as a function of $\theta_1$ and $\theta_2$.  The result is shown in figure \ref{fig:esurf}.  The energy landscape with the given hardware general ansatz is shown as a function of the rotation angles $\theta_1$ and $\theta_2$ of the $R_y$ and $R_z$ operators respectively.  The global minimum is located at $E\simeq-1.75 $MeV at $(\theta_1,\theta_2)\simeq(3\pi/16,0)$.  Since $\theta_2=0$ a suitable problem-specific ansatz could be used without the $R_z$ gate.

For the N=4 case of $H_N$, which is beyond the size attempted on a real quantum computer in previous work with the JW encoding\cite{dumitrescu_cloud_2018}, we use an ansatz
\vskip4mm
\centerline{ \begin{quantikz}
			                     \lstick{\ket{0}} & \qw & \gate{R_y(\theta_0)} & \qw & \ctrl{1} & \qw & \gate{R_y(\theta_2)} & \qw & \ctrl{1} &\qw \\
			                    \lstick{\ket{0}} & \qw & \gate{R_y(\theta_1)} & \qw & \targ{} & \qw &  \gate{R_y(\theta_3)} & \qw & \targ{} & \qw\\ 
			        \end{quantikz}}
	
	\noindent	on the IBM\cite{IBMQ} Manilla quantum computer\footnote{\hyperlink{https://quantum-computing.ibm.com/services?services=systems&system=ibmq_manila}{https://quantum-computing.ibm.com/services?services=systems\&system=ibmq\_manila}}, with zero-CNOT extrapolation error mitigation \cite{Ying_Li_2017}, to obtain a value of the ground state energy via the VQE algorithm of $ -2.108$ MeV $\pm 0.080$ MeV, equal within errorbars to the exact value of $-2.144$ MeV.

\section{Minimization of Variance}
In the standard VQE\cite{mcclean_theory_2016,cerezo_variational_2021} one invokes the variational principle from quantum mechanics which states that the expectation value of a Hamiltonian with an arbitrary wave function will have a value equal or higher to the true ground state energy.  Thus by producing a wave function ansatz with one or more variable parameters, one can minimize the expectation value with respect to the parameters and obtain an approximation to the true ground state.  

If, instead of the expectation value of $H$, we calculate the variance
\begin{equation}
    \sigma^2 = \langle H^2\rangle - \langle H\rangle^2
\end{equation}
and minimize this quantity with respect to parameters in the wave function ansatz, we may be able to locate any eigenstate of the Hamiltonian, whether the ground state or excited state.  This method, and adaptations of it, have been applied in quantum chemistry\cite{arxiv.2006.15781,arxiv.2111.05176}.

To see the method in operation, we apply it to the deuteron problem as described in section \ref{sec:encoding}, with a one-parameter ansatz of the form used suitable for a JW encoding of the $N=2$ Hamiltonian\cite{dumitrescu_cloud_2018}
\vskip4mm
			       \centerline{\begin{quantikz}
				\lstick{\ket{0}} & \qw & \gate{X}
                     & \targ{} & \qw\\
                \lstick{\ket{0}} & \qw & \gate{R_y(\theta)} & \ctrl{-1}&\qw \\
				\end{quantikz}}
				
\begin{figure}
    \centering
    \includegraphics[width=0.7\textwidth]{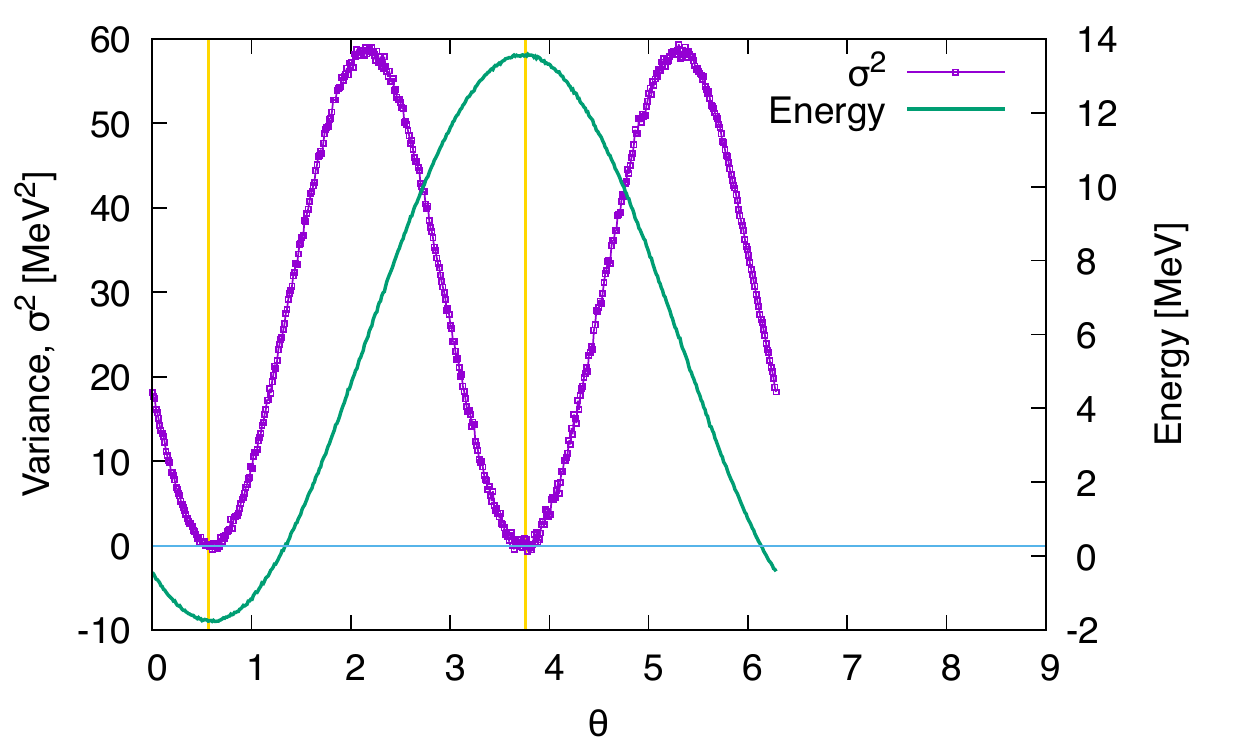}
    \caption{Variance ($\sigma^2$) of Hamiltonian with parameterised wave function ansatz, and energy for same parameterised wave functions.  Golden vertical lines are drawn at the minima of the variance to guide the eye to the corresponding energy values.}
    \label{fig:my_label}
\end{figure}

With a single parameter it is practical to perform a parameter sweep to visualise the dependence of Hamiltonian variance on the variational parameter, which is shown in figure \ref{fig:my_label}.

These calculations were performed using the QASM backend of the IBM qiskit environment \cite{Qiskit}.  One sees here that there are two minima in the variance with $\sigma^2=0$.  One of these corresponds to the global minimum of the energy - hence the ground state.  The other minimum variance is at $E=13.586$ MeV, close to the analytic value of the second eigenstate of $E=13.563$ MeV.

Given the apparent success of the method, we further explore variance minimization with the VQE algorithm by exploring the LMG model\cite{lipkin_validity_1965}.  This model is a simplified version of the nuclear shell model in which two single-particle levels, separated by energy $\varepsilon$, are each $N$-fold degenerate, with a total of $N$ fermionic particles in the system.  The non-interacting ground state then consists of all particles being in the lower level.  A two-body interaction term gives a probability of scattering pairs of particles into the upper level, and a rich set of behaviour can be seen depending on choice of number of particles, $N$, and interaction strength, such as the analogue of a nuclear shape transition\cite{agassi_validity_1966}, or phase transitions in the thermodynamic limit\cite{arxiv.1102.1583}.  It is a standard test system in nuclear physics, and as noted in the introduction has already been explored in the context of quantum computing algorithms\cite{cervia_lipkin_2021,robbins_benchmarking_2021}.
 
 We consider here the $N=2$ and $N=4$ LMG model Hamiltonians as implemented with the reduced qubit representation of section \ref{sec:encoding} and the hardware-general ansatzes shown there.  
 
 Using IBM's qiskit system and it \textit{statevector simulator} backend - essentially an exact quantum-mechanical calculation, alongside a standard VQE algorithm to seek the minimum variance, we perform multiple VQE simulations with randomized starting points, to see which local minimum in the variance is found.  For each zero variance, the corresponding energy is recorded.  Running many times, we find we are able to map out the exact series of energy levels and obtain the full spectrum, consisting of two levels in the $N=2$ case and four levels in the $N=4$ case.  
 
\begin{figure}
    \centering
    \includegraphics[width=0.45\textwidth]{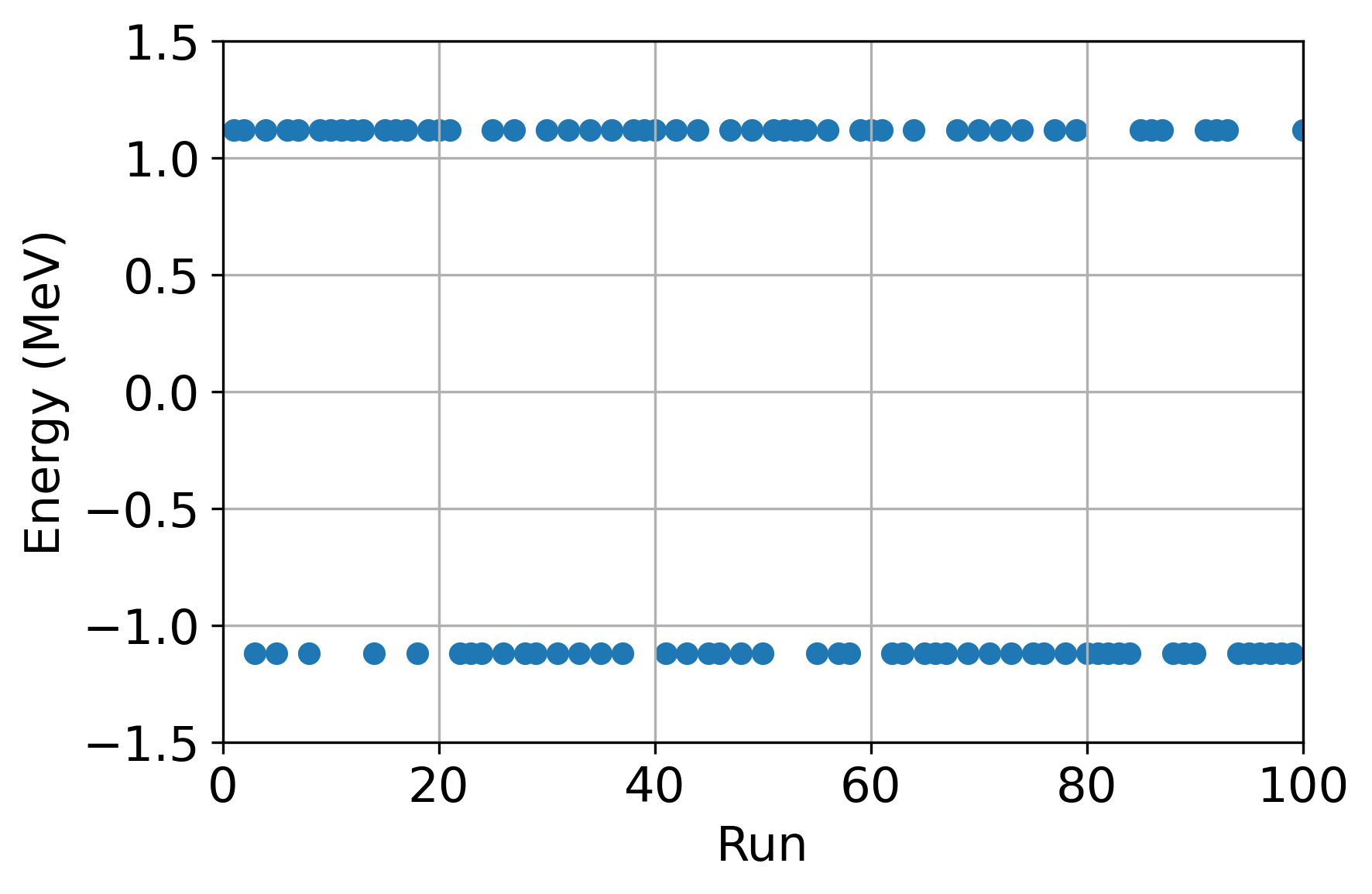}\includegraphics[width=0.45\textwidth]{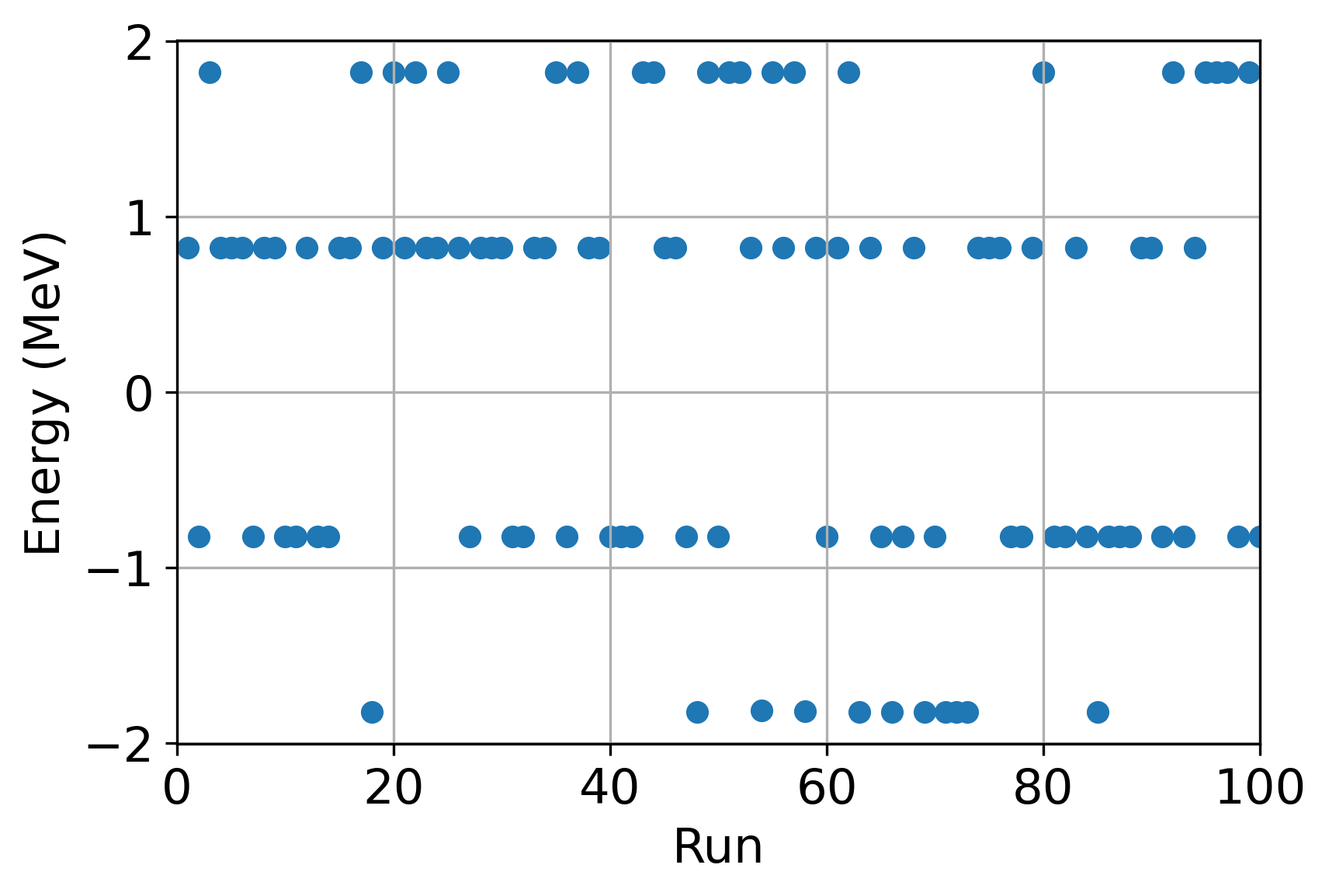}
    \caption{Energies for each VQE run (labelled on x-axis) for VQE calculation in which Hamiltonian variance is minimized for the N=2 LMG model (left plot) and N=4 LMG model (right plot).  In both calculations, the interaction strengths in the notation of the original LMG paper\cite{lipkin_validity_1965} is $V/\epsilon = 0.5$ and $W=0$.}
    \label{fig:lipkin_vqe}
\end{figure} 
Investigations of this method on cloud quantum computers, with associated errors, and resulting error analysis mitigation methods, are underway.

\section{Conclusions}
We have presented results of VQE calculatons on real and simulated quantum computers.  The feasibility of using a reduced qubit encoding compared to the Jordan-Wigner mapping has been explored, and found to give good results, though the wave function ansatz may be more complicated in some cases.  A method in which the Hamiltonian variance, rather than the energy, is minimized has been shown in principle to work on a simulated quantum computer, as well as through a standard quantum mechanical calculation. 

\acknowledgments 
 
This work was funded by AWE. We acknowledge the use of IBM Quantum services for this work. The views expressed are those of the authors, and do not reflect the official policy or position of IBM or the IBM Quantum team. In this paper we used ibmq\_manilla, which is one of the IBM Quantum Falcon Processors. UK Ministry of Defence \copyright  Crown owned copyright 2022/AWE.

\bibliography{pds} 
\bibliographystyle{spiebib} 

\end{document}